\documentstyle[sprocl,epsfig]{article}

\def\be{\begin{equation}}
\def\ee{\end{equation}}
\def\bea{\begin{eqnarray}}
\def\eea{\end{eqnarray}}

\def \ben{\begin{enumerate}}
\def \een{\end{enumerate}}
\def \bit{\begin{itemize}}
\def \eit{\end{itemize}}

\def \branch{{\cal B}}
\def \eff{\hbox{eff}}
\def \gev{{\hbox{GeV}}}

\def \cl#1{{#1\%\ \mathrm{C.L.}}}

\def \fig#1{Fig.~\ref{#1}}

\def \nn{\nonumber}

\def \D{\Delta}
\def \g{\gamma}

\def \d{\delta}
\def \e{\epsilon}

\def \m{\mu}

\def \o{\omega}

\def \s{\sigma}

\def \alphas{\alpha_s}

\def\o{{\cal O}}

\begin{document}
\begin{flushright}
DESY 01-224 \\
December 2001\\
\end{flushright}

\vspace*{1.5cm}
\begin{center}
{\Large \bf
\centerline{Signatures of Supersymmetry in $B$ Decays - A Theoretical
Perspective}}
\vspace*{1.5cm}
{\large A.~Ali}
\vskip0.2cm
Deutsches Elektronen-Synchrotron DESY, Hamburg \\
Notkestra\ss e 85, D-22603 Hamburg, FRG\\

\vspace*{6.0cm}
{\it {\bf
Invited Talk; to be published in the Proceedings of the International
Conference on Flavor Physics ICFP2001, Zhang-Jia-Jie City, Hunan Province,
Peoples Republic of China, May 31 -- June 6, 2001}}.

\end{center}
\thispagestyle{empty}

\newpage
\setcounter{page}{1}
\title{Signatures of Supersymmetry in $B$ Decays - A Theoretical
Perspective}

\author{A.~ALI}

\address{Deutsches Elektronen-Synchrotron DESY, Notkestra\ss e 85\\ 
 D-22603 Hamburg, Germany\\E-mail: ali@mail.desy.de} 

\maketitle\abstracts{We discuss precision tests of the standard 
model in radiative and semileptonic rare $B$-decays and CP-violating 
asymmetries, and possible signatures of supersymmetry 
in these processes. Motivated by current data,
and with an eye on the forthcoming measurements, we restrict ourselves
to the processes in which the bench marks set by the standard model have
either been met by ongoing experiments or are expected to be met shortly.
This includes the mixing-induced CP asymmetry, measured through $\sin 2
\beta$, and branching ratios for the radiative and semileptonic rare
decays $B \to X_s \gamma$, $B \to K^* \gamma$,
$B \to X_s \ell^+ \ell^-$ and $B \to (K,K^*) \ell^+ \ell^-$, where
$\ell^\pm =e^\pm, \mu^\pm$.}

\section{Introduction}
 Experimental $B$ physics is making big strides,
thanks largely to the superb performance of the
KEK and SLAC $B$ factories and the BELLE and BABAR detectors, but
also due to the solid foundation provided by
CLEO, the LEP and SLC experiments, and by the CDF collaboration. 
The rich harvest of new experimental results includes
the first measurements of CP violation 
in the $B$-meson system $B_d^0$ and $\overline{B_d^0}$, involving 
statistically significant results by the BABAR~\cite{Aubert:2001cp} and 
BELLE~\cite{Abashian:2001cp}
collaborations. The CP asymmetry in question, $a_{J\psi
K_s}$, is  measured through the time-dependent 
difference in the decay rates for $B_d^0 \to J/\psi K_s$ and its
CP-conjugate process $\overline{B_d^0} \to J/\psi K_s$ (as well as a
number of other related final states such as $\psi(2S) K_s$, $J/\psi K_L$
etc.). Data on this asymmetry, allowing a clean determination 
of $\sin 2 \beta$ (where $\beta$ is one of the angles of the unitarity 
triangle UT), is now quantitative and can be 
meaningfully compared  with the standard model (SM) predictions of the 
same. Likewise, it can be used to put  
constraints on additional CP-violating phases in 
beyond-the-SM (BSM) scenarios and
we will illustrate this in terms of an additional phase, called 
$\theta_d$, in the context of  a supersymmetric theory.

Equally interesting
from the point of view of precision tests of the SM and searches of BSM 
physics are a number of
flavour-changing-neutral-current (FCNC) processes, of which the decay
$b \to s \gamma$ is so far the most 
significant~\cite{Alam:1995aw,cleobsg,alephbsg,bellebsg}. 
Here also, SM has not only survived a rather crucial experimental test
involving quantum (loop) effects in the FCNC sector, it
has done so with a comfortable
ease~\cite{Chetyrkin:1997vx,Kagan:1999ym,Gambino:2001ew}.
Experiments at the B 
factories are about to cross (or, have probably already crossed) the next 
milestone 
in rare $B$-decays, involving semileptonic FCNC decays. We have in mind 
here the decays $B \to (X_s, K^*,K) \ell^+ \ell^-$, where the current 
experimental limits~\cite{bellebsll,Abe:2001dh,Aubert:2002aw} are fast 
approaching the SM-sensitivity~\cite{Ali:1996bm,Ali:2000mm,Ali:2001jg}. In 
fact in the decays $B \to K \ell^+ \ell^-$, $\ell=e,\mu$, first 
measurements by the
BELLE collaboration~\cite{Abe:2001dh} are at hand, though the BABAR
collaboration~\cite{Aubert:2002aw} still does not
see any signal in this channel. So, more data is needed, which luckily is
on its way. The next experimental milestone in this 
field will be reached with the
measurement of the inclusive decays $B \to X_s \ell^+ 
\ell^-$, since the SM-theory is now quantitative, in particular in the 
low dilepton invariant mass region~\cite{Ali:2001jg,BMU,AAGW}. Data can be 
used to extract effective Wilson coefficients in a general theoretical
scenario which we shall discuss, drawing heavily from a recent 
analysis~\cite{Ali:2001jg}.   

\section{Constraints on $\sin 2 
\beta$ and an additional CP-violating phase $\theta_d$}
\label{sec:sin2beta}

 As is by now folk-lore, the measurement of $a_{J\psi K_s}$ yields  $\sin 
2 \beta$ in the SM. The
current experimental value of this quantity (including a 
scale factor in conformity with the Particle Data Group prescription) is 
$\sin 2 \beta =0.79 \pm
0.12$, and is now dominated by the BABAR~\cite{Aubert:2001cp}
 ($\sin 2 \beta =0.59 \pm 0.14 ({\rm stat}) \pm 0.05 ({\rm syst})$),
and  the BELLE~\cite{Abashian:2001cp}
($\sin 2 \beta =0.99 \pm 0.14 ({\rm stat}) \pm 0.06 ({\rm syst})$)
measurements.
 SM-predictions of $\sin 2 \beta$ based on indirect measurements
of the sides of the unitarity triangle lie in the
range $\sin 2 \beta = 0.6$ -- $0.8$~\cite{ali1}. From this,
we tentatively conclude that the current experiments and SM
are in reasonable agreement with each other in $\sin 2 \beta$.
However, this {\it rapport} will be tested very precisely in future
and it is sensible to estimate the magnitude of a BSM-phase allowed
by current data.  

 In popular extensions of the SM, such as the minimal supersymmetric
standard model (MSSM), one anticipates supersymmetric contributions to
FCNC processes, in particular $\D M_{B_d}$, $\D M_{B_s}$ (the mass
differences in the $B_d^0$ -$\overline{B_d^0}$ and $B_s^0$ 
-$\overline{B_s^0}$ systems), and
$\epsilon_K$, characterizing the mixing-induced CP-asymmetry ${\cal 
A}_{\rm CP}^{\rm mix}$ in the
$K^0$ -$\overline{K^0}$ system. However, if the 
Cabibbo-Kobayashi-Maskawa (CKM) matrix remains
effectively the only flavour changing (FC) structure, which is the
case if the quark and squark mass matrices can be simultaneously
diagonalized, and all other FC interactions
are associated with rather high scales, then all hadronic flavour
transitions can be interpreted in terms of the same unitarity  
triangles which one encounters in the SM.  In particular, in these
theories $a_{\psi K_s}$ measures the same quantity $\sin 2 \beta$ as
in the SM. These models, usually called the minimal flavour
violating (MFV) models~\cite{giudice}, are structured so that
the SUSY contributions to $\D M_{B_d}$, $\D
M_{B_s}$, and $\epsilon_K$ have the same CKM-dependence as the SM top
quark contributions in the box diagrams (denoted below by
$C_1^{Wtt}$).  Hence, supersymmetric effects for the UT-analysis
can be effectively incorporated in terms of a single common parameter 
$f$ by the following replacement \cite{ali2}:
\be
\e_K, \; \D M_{B_s} \; \D M_{B_d}, \; a_{\psi K_S}:
 C_1^{Wtt} \rightarrow  C_1^{Wtt} (1+f) ~.
\label{fmfv}
\ee
The parameter $f$ is positive definite and real, implying that there are
no new phases in any of the quantities specified above. The size of $f$
depends on the parameters of the supersymmetric models. Given a value of
$f$, the CKM unitarity fits can be performed in these scenarios much 
the same way as they are  done for the SM. Qualitatively, the CKM-fits
in MFV models yield the following pattern for the three inner angles of
the UT:
\be
\beta^{\rm MFV} \simeq \beta^{\rm SM}~; \; ~~\gamma^{\rm MFV} <
\gamma^{\rm SM}~; \; ~~\alpha^{\rm MFV} > \alpha^{\rm SM}~,
\label{mvfcp}
\ee
and a recent CKM-fit along these lines yields the following
central values for the three angles \cite{ali1}:
\bea
f &=& 0 ~({\rm SM}): ~~(\alpha, \beta, \gamma)_{\rm central} = (95^\circ,
22^\circ,63^\circ)~, \nn\\
f &=& 0.4 ~({\rm MFV}): ~(\alpha, \beta, \gamma)_{\rm central} =
(112^\circ, 20^\circ,48^\circ)~.
\label{mfvsmcent}
\eea
leading to $(\sin 2\beta)^{\rm SM}_{\rm central} \simeq 0.70$ and
$(\sin 2\beta)^{\rm MFV}_{\rm central} \simeq 0.64$.
Thus, what concerns $\sin 2 \beta$, the SM and the MFV models
give similar results from the UT-fits, unless much larger values for the
parameter $f$ are admitted which are now unlikely
due to the existing constraints on the MFV-SUSY parameters.
 
However, in a  general extension of the SM, one expects that
all the quantities appearing on the l.h.s. in Eq.~(\ref{fmfv})
will receive independent additional contributions. In this case, the
magnitude and the phase of the off-diagonal elements in the $B_d^0$
-$\overline{B_d^0}$ and $B_s^0$ - $\overline{B_s^0}$ mass matrices
can be parametrized as follows \cite{nelson,wolf-silva}:
\bea
M_{12}(B_d) &=& {\langle \bar B_d | H_{eff}^{\D
B=2} |B_d \rangle \over 2 M_{B_d}} = r_d^2 e^{2 i \theta_d}
M_{12}^{SM}(B_d)~,
\nn \\
M_{12}(B_s) &=& {\langle \bar B_s | H_{eff}^{\D
B=2} |B_s \rangle \over 2 M_{B_s}} = r_s^2 e^{2 i  \theta_s}
M_{12}^{SM}(B_s)~.
\label{mbdgen}
\eea
where $r_d$ ($r_s$) and $\theta_d$ ( $\theta_d$) characterize,
respectively, the magnitude and the phase of the new physics
contribution to the mass difference $\D M_{B_d}$ ($\D M_{B_s}$). It
follows that a measurement of $a_{\psi K_s}$ would not measure $\sin 2
\beta$, but rather a combination $\sin 2 (\beta +
\theta_d)$. In this scenario,
one also expects new contributions in $M_{12}(K^0)$, bringing in their
wake additional parameters ($r_\epsilon$, $\theta_\epsilon$). They  
will alter the profile of CP-violation in the decays of the neutral
kaons.

It is obvious that in such a general theoretical scenario, which
introduces six {\it a priori} independent parameters, the predictive
power vested in the CKM-UT analysis is lost. If the idea is to retain
this predictivity, at least partially, then one has to work within a
more limited framework. A model along these lines was
introduced using the language of minimal
insertion approximation (MIA) \cite{mia,mia2} in a supersymmetric context.
In this model~\cite{Ali:2001ej}, all FC
transitions which are not generated by the CKM mixing matrix are
proportional to the properly normalized off--diagonal elements of the
squark mass matrices:
\be
(\d_{ij})^{U,D}_{AB} \equiv {(M^2_{ij})^{U,D}_{AB} \over
M_{\tilde q_i} M_{\tilde q_j}}
\label{miadef}
\ee
where $i,j=1,2,3$ and $A,B=L,R$.
The dominant effect of the
non-CKM structure contained in the MIA-parameters is to    
influence mainly the $b \to d$ and $s\to d $ transitions while the $b
\to s$ transition is governed by the MFV-SUSY and the SM contributions
alone.  For what concerns the quantities entering in the UT analysis,
the following pattern for the supersymmetric contributions emerges in
this model:
\bea
\label{f}
\D M_{B_s}: & C_1^{Wtt} \rightarrow & C_1^{Wtt} (1+f) \\
\e_K, \; \D M_{B_d}, \; a_{\psi K_S}: & C_1^{Wtt} \rightarrow &
         C_1^{Wtt} (1+f)+C_1^{MI} \equiv
         C_1^{Wtt} \left(1+f+ g \right)
\label{g}
\eea
where the parameters $f$ and $g=g_R + i g_I$ represent normalized   
(w.r.t. the SM top quark $Wtt$) contributions from the MFV and MIA
sectors, respectively.  Hence, in the UT-analysis the contribution from
the supersymmetric sector can be parametrized by two real parameters
$f$ and $g_R$ and a parameter $g_I$, generating a phase $\theta_d$,
which is in general non-zero due to the complex nature of the
appropriate mass insertion parameter. A
precise measurement of $a_{\psi K_s}$ would fix this argument
($=\theta_d$).

\begin{figure}[H]
\begin{center}
\epsfig{file=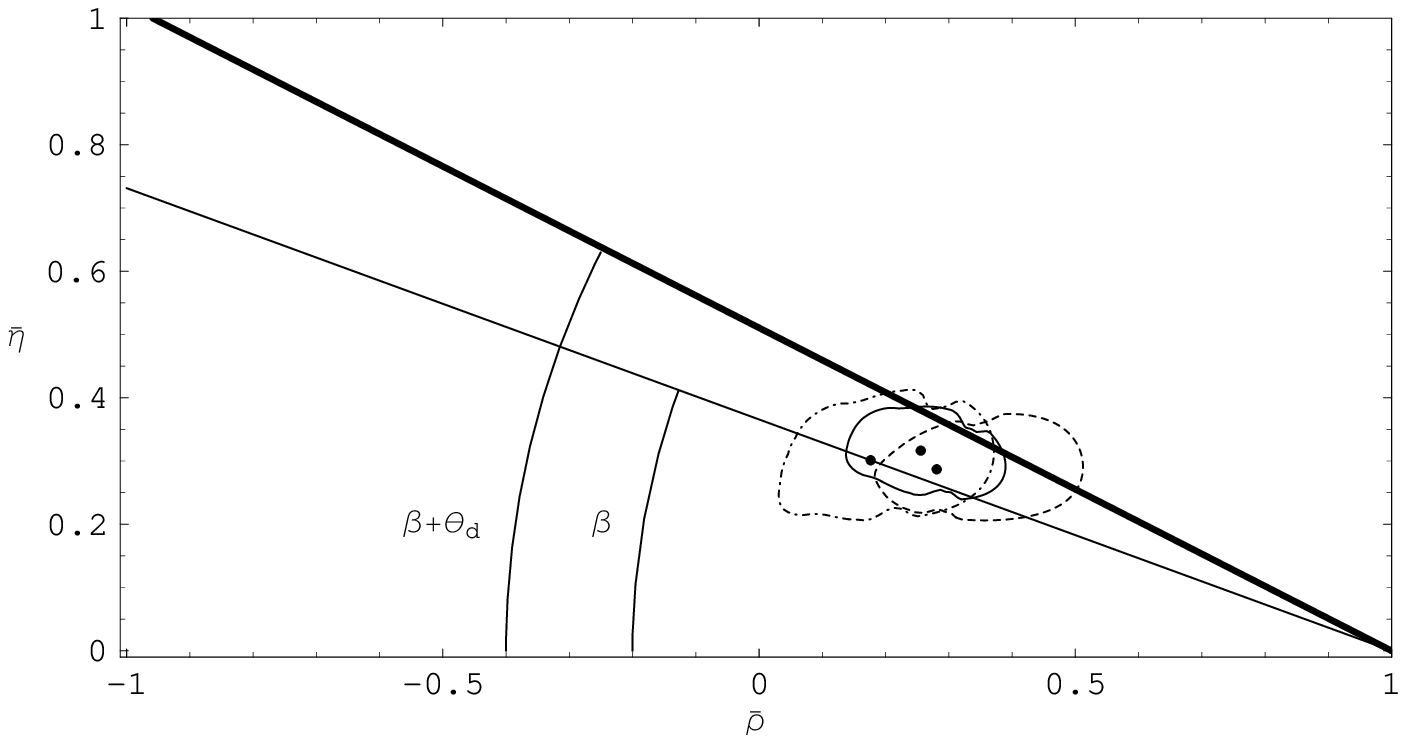,width=0.9\linewidth}
\caption{\it Allowed $95 \; \% \; C.L.$ contours in the
$(\bar\rho,\bar\eta)$
plane.  The solid contour corresponds to the SM case, the dashed contour
to the Minimal Flavour Violation case with $(f=0.4, \; g=0)$ and the
dashed--dotted contour to the Extended-MFV model discussed in
the text $(f=0, \; g_R=-0.2, \; g_I = 0.2)$. (From 
Ref.~25.)}
\label{cont}
\end{center}
\end{figure}

\begin{figure}[H]
\begin{center}
\epsfig{file=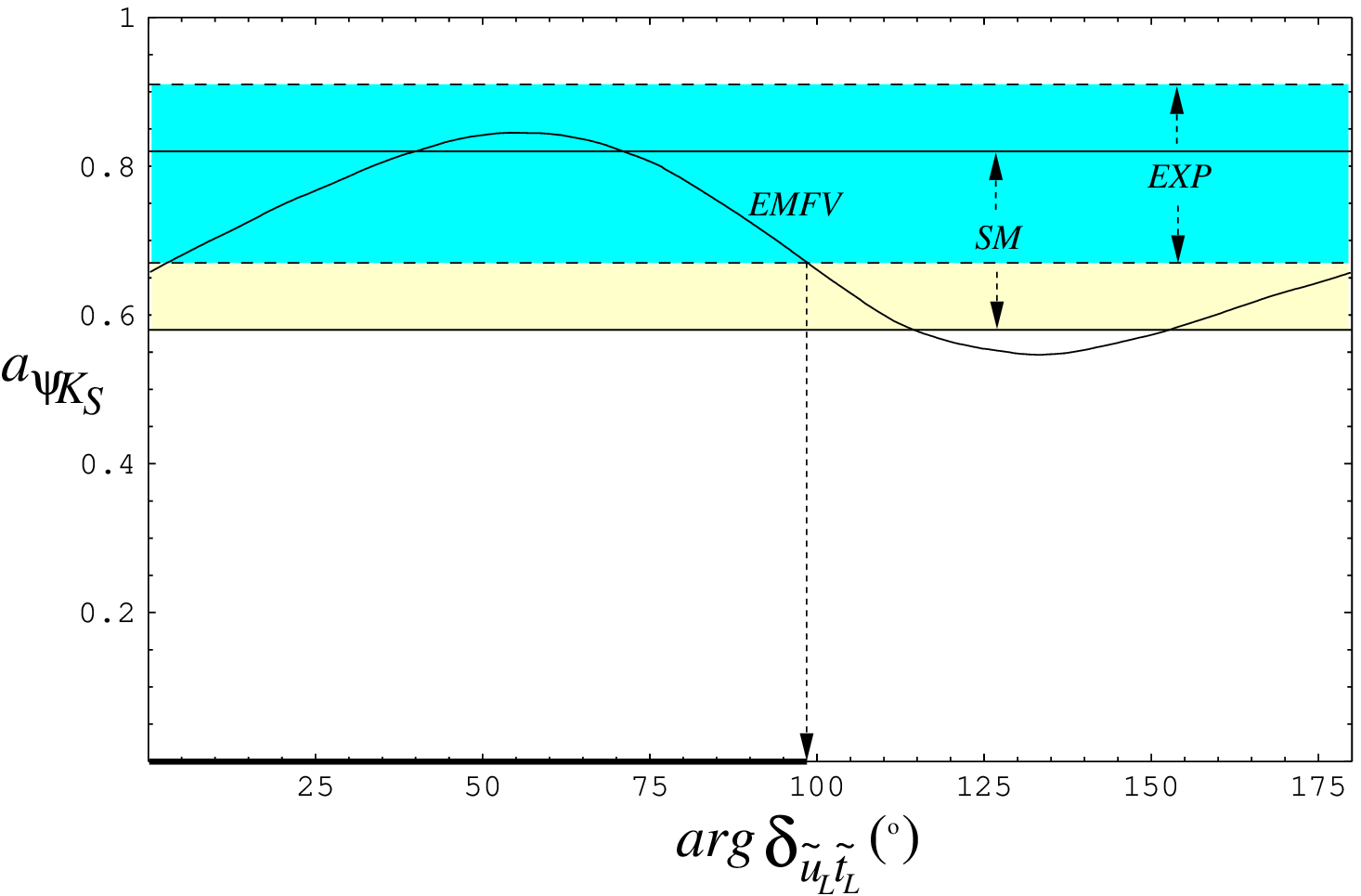,width=0.495\linewidth}
\epsfig{file=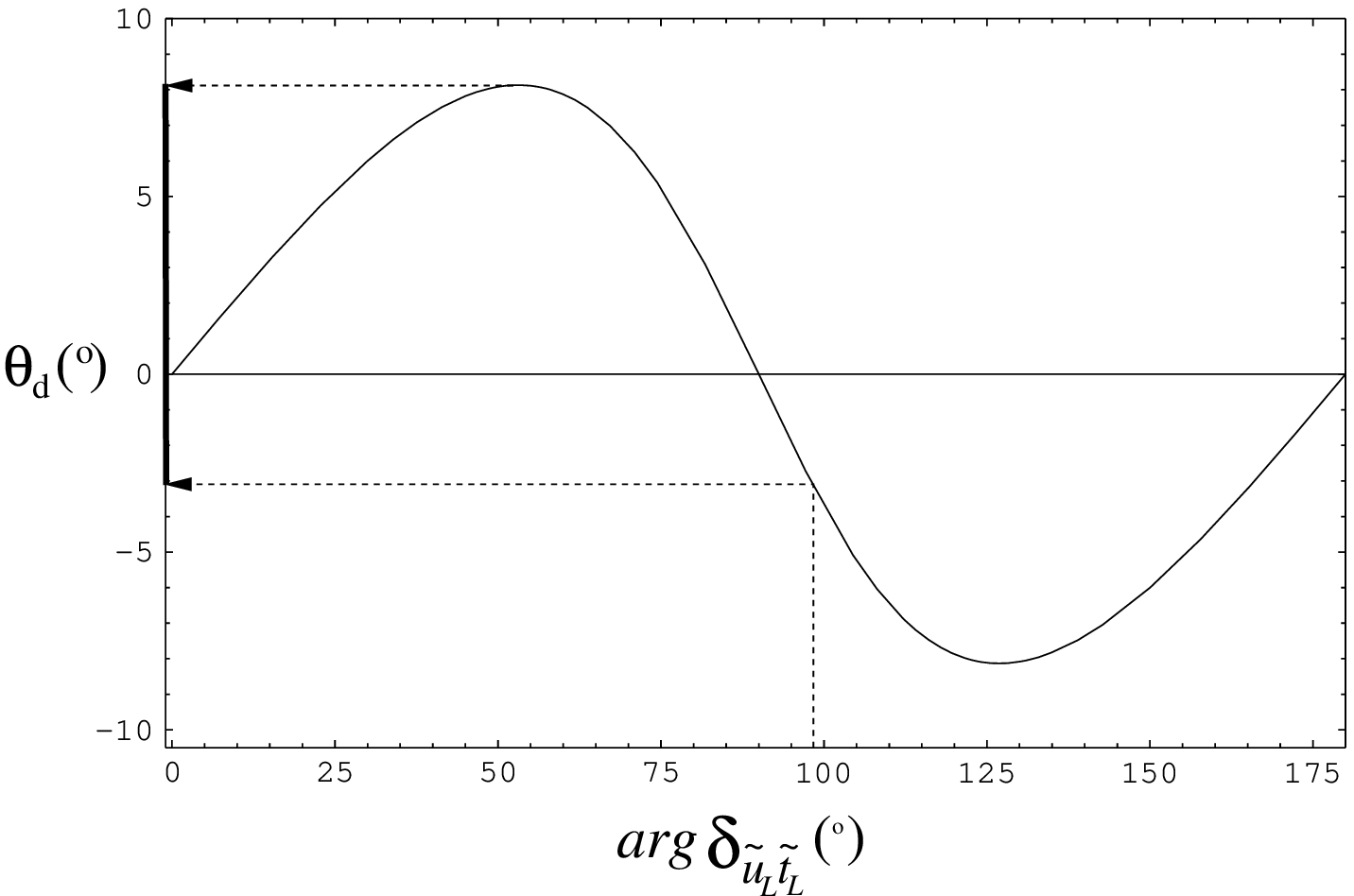,width=0.495\linewidth}
\caption{\it The $CP$ asymmetry $a_{\psi K_S}$ as a function of $\arg
\delta_{\tilde u_L \tilde t_2}$ expressed in degrees. The solid curve
corresponds to the Extended-MFV model $(f=0, \;|g|=0.28)$. The light
and dark shaded bands correspond, respectively, to the allowed $1\;\s$
region in the SM ($0.58 \leq a_{\psi K_S} \leq 0.82$) and the current
$1 \;\s$ experimental band ($0.67 \leq a_{\psi K_S} \leq 0.91$). The
plot on the right shows the correlation between $\arg \delta_{\tilde
u_L \tilde t_2}$ and the angle $\theta_d$: $\theta_d ={1\over 2}\arg
(1+f+ |g| e^{2 i \arg \delta_{\tilde u_L \tilde t_2}}),\; (\hbox{mod}
\; \pi)$. The experimentally allowed region flavours $0^\circ < \arg
\delta_{\tilde u_L \tilde t_2}< 100^\circ$ that translates into $
-3^\circ < \theta_d < 8^\circ$. (From Ref.~25.)}
\label{apsiks}
\end{center}
\end{figure}

 The impact of the Extended-MFV model on the profile of the
unitarity triangle in the $(\bar\rho,\bar\eta)$ plane is shown in 
Fig.~\ref{cont}, which also shows the
corresponding profiles in the SM and MFV models: 
the solid contour corresponds to the SM $\cl{95}$, the dashed one to a
typical MFV case ($f=0.4$, $g= 0$) and the dotted--dashed one to an
allowed point in the Extended-MFV model ($f=0$, $g_R=-0.2$ and
$g_I=0.2$). All three models give comparable fits. In \fig{apsiks}
the $CP$ asymmetry $a_{\psi K_S}$ is plotted as a function of $\arg
\delta_{\tilde u_L \tilde t_2}$ (expressed in degrees).
Here, $\delta_{\tilde u_L \tilde t_2}$ is a linear combination~\cite{mia2} 
of
$(\d_{13})^{U}_{LR}$ and $(\d_{13})^{U}_{LL}$. The light and
dark shaded bands correspond, respectively, to the SM and the    
experimental $1\;\s$ allowed regions. The solid line is drawn for
$f=0$ and $|g| =0.28$.  The experimental band flavours $\arg
\delta_{\tilde u_L \tilde t_2}$ in the range
$[0^\circ,100^\circ]$. Employing the explicit dependence
\be
\theta_d ={1\over 2}\arg (1+f+ |g| e^{2 i \arg \delta_{\tilde u_L  
\tilde t_2}}) \;\;\; (\hbox{mod} \; \pi)\; ,
\ee
the above phase interval is translated into
\begin{equation}
 -3^\circ < \theta_d < 8^\circ \; ,
\label{thetad}
\end{equation}
for the assumed values of $|g|$ and f, which is a typical
range for $\theta_d$ for the small angle solution with the current
values of $a_{\psi K_S}$.

 Such a non-zero angle would have measurable
consequences in $b \to d$ transitions~\cite{Ali:2001ej,Ali:2000zu}, such 
as the isospin-violating ratio
$\Delta^{\pm 0} =
\frac{\Gamma (B^\pm \to \rho^\pm \gamma)}
     {2 \Gamma (B^0 (\bar B^0)\to \rho^0 \gamma)} - 1$,
and in direct CP-violating asymmetries  
${\cal A}_{\rm CP} (\rho^\pm \gamma) =
\frac{{\cal B} (B^- \to \rho^- \gamma) - {\cal B} (B^+ \to \rho^+ \gamma)}
     {{\cal B} (B^- \to \rho^- \gamma) + {\cal B} (B^+ \to \rho^+ 
\gamma)}$, which may deviate measurably from their SM ranges.
Recently, these quantities have been calculated in the SM by taking into 
account explicit $O(\alpha_s)$ corrections~\cite{Bosch:2001gv,Ali:2001ez},
using the so-called Large-Energy-Effective-Theory (LEET) 
approach~\cite{Charles:1999dr}. While there is considerable parametric 
uncertainty in the determination of ${\cal A}_{\rm CP} (\rho^\pm \gamma)$
(and its analogue ${\cal A}_{\rm CP} (\rho^0 \gamma)$), the estimates of
$\Delta^{\pm 0}$ are quantitative. Testing these predictions should be 
feasible at the B factories in the next several years.

\section{Inclusive Decay rate for $B \to X_s \gamma$ in the
SM and SUSY}\label{sec:incl-rad}
The effective Hamiltonian in the SM inducing the $b\to s\g$ transitions,
obtained by integrating out the heavier degrees of freedom,
can be expressed as follows:   
\begin{eqnarray}
    \label{Heff}
    {\cal H}_{\eff} =  - \frac{4G_F}{\sqrt{2}} V_{ts}^* V_{tb}
    \sum_{i=1}^{8} C_i(\mu) \, O_i (\mu)\quad ,
\end{eqnarray}
where $G_F$ is the Fermi coupling constant and the CKM dependence has 
been made explicit; $O_i(\mu)$ are dimension-six operators at the scale 
$\mu$, and $C_i(\mu)$ are the corresponding Wilson coefficients.
Of these, the dominant operators are
${\cal O}_1 \sim (\bar{s}_L\gamma_\mu T^a q_L)(\bar{q}_L\gamma^\mu
T^a b_L)$,
${\cal O}_2 \sim (\bar{s}_L\gamma_\mu q_L)(\bar{q}_L\gamma^\mu b_L)$,
and the magnetic moment operators ${\cal O}_7 \sim (\bar{s}_L \sigma_{\mu
\nu}
b_R)F^{\mu \nu}$ and ${\cal O}_8 \sim (\bar{s}_L
\sigma_{\mu \nu} b_R)T^a G^{a,\mu \nu}$, where $F^{\mu \nu} (G^{a,\mu
\nu})$ is the electromagnetic (chromomagnetic) field strength tensor, with
$T^a (a=1,...,8)$ being the SU(3) group generators. Current theoretical
precision of the $b\to s\g$ decay rate is limited to 
$\o (\alphas)$, consisting of the anomalous
dimension matrix in the next-to-leading order (NLO), the commensurate 
matching
conditions, and the virtual and bremsstrahlung contributions. Also, the
leading power corrections in $1/m_b$ and $1/m_c$ have been calculated.
 The present experimental average of the branching   
ratio~\cite{Alam:1995aw,cleobsg,alephbsg,bellebsg}
\be
 {\cal B}(B \to X_s \gamma)= (3.22 \pm 0.40)\times 10^{-4}~,
\label{eq:bsgamexp}
\ee
 is
in good agreement with the next-to-leading order prediction
of the same in the SM, estimated 
as \cite{Chetyrkin:1997vx,Kagan:1999ym} ${\cal B}(B \to X_s
\gamma)_{\rm SM}= (3.35 \pm 0.30)\times 10^{-4}$
for the pole quark mass ratio
$m_c/m_b=0.29 \pm 0.02$, rising to \cite{Gambino:2001ew} ${\cal B}(B \to 
X_s \gamma)_{\rm SM}= (3.73 \pm 0.30)\times 10^{-4}$, if one uses
instead the input value $m_c^{\overline{\rm MS}}(\mu)/m_b^{\rm pole}=0.22 
\pm 0.04$, where $m_c^{\overline{\rm MS}}(\mu)$ is the charm quark mass in
the $\overline{\rm MS}$-scheme, evaluated at a scale $\mu$ in the range
$m_c < \mu < m_b$. The inherent uncertainty reflects the present
accuracy of the theoretical branching ratio
 and the imprecise knowledge of the quark masses,
in particular $m_c$ and $m_b$. Precise measurements of the
photon energy spectrum in $B \to X_s \gamma$ decays may
help in decreasing some of these uncertainties.

 The agreement between
experiment and the SM for the $B \to X_s \g$ decay rate is quite 
impressive and this has been used to put
non-trivial constraints on the BSM-physics, in particular
supersymmetry~\cite{Hewett:1996ct,Kagan:1999ym}. A recent
analysis~\cite{Ali:2001jg} 
along these lines is discussed here for illustration. Following earlier 
works, the integrated $B\to X_s \g$ branching ratio can be solved as a 
function of the quantities $R_{7,8}(\mu_W)\equiv
C_{7,8}^{\rm tot} (\mu_W) / C_{7,8}^{\rm SM} (\mu_W)$,
where $R_{7,8}^{\rm tot}=R_{7,8}^{NP} + R_{7,8}^{SM}$. Taking the scale 
$\m_W=M_W$, for the purpose of the renormalization group evolution (RGE)  
also for the supersymmetric contributions, and  imposing
the experimental bound ${\cal B}(B \to X_s \g)=(3.22 \pm 0.40) \times
10^{-4}$, the corresponding allowed regions in the 
$[R_7(\mu_W),R_8(\mu_W)]$ plane are worked out.
Evolving the allowed regions to the scale $\mu_b=2.5 \; \gev$
and assuming that new physics only enters in the effective Wilson
coefficients $C_{7,8}$, the corresponding low--scale
bounds in the plane $[R_7(2.5 \; \gev),R_8(2.5 \; \gev)]$ are obtained,
yielding
\bea
\label{a7neg}   
 0.785 \leq R_7 (2.5\;\gev) \leq 1.255 &\Rightarrow&
        -0.414 \leq C_7^{\rm tot,<0} (2.5 \; \gev) \leq -0.259 \; ,
\nonumber\\
 -1.555 \leq R_7 (2.5\;\gev) \leq -1.200 &\Rightarrow&
        0.396 \leq C_7^{\rm tot,>0} (2.5 \; \gev) \leq 0.513 \; .  
\label{a7pos}
\eea
Depending on the sign of $C_7^{\rm eff}$, there are two allowed solutions
- called $C_7^{\rm tot}$--positive and $C_7^{\rm tot}$--negative  
solutions. SM corresponds to the point $(R_7 (\m), R_8(\m))$ $=(1,1)$.
Flavour-blind supersymmetric theories, such as SUGRA, allow points in the 
vicinity of
the SM, though in a more general supersymmetric scenario, both 
$C_7^{\rm tot}$--positive and $C_7^{\rm tot}$--negative
solutions are allowed~\cite{LMSS}. These two scenarios can be 
distinguished, in  principle, by measurements of the decays $B \to 
(X_s,K,K^*) \ell^+ \ell^-$, which we discuss next.  

\section{Inclusive Decays $B \to X_s \ell^+ \ell^-$ in the
SM and SUSY}\label{sec:incl-excl-sl}

  The inclusive decays $B \to X_s \ell^+ \ell^-$ and the
corresponding exclusive decays such as $B \to (K,K^*) \ell^+ \ell^-$ allow 
to get more detailed information on the flavour structure of the SM,
and hence offer new search strategies for BSM physics.
The effective Hamiltonian governing these decays in the SM is obtained
by enlarging the sum given in Eq.~(\ref{Heff}) by the addition of two
more terms involving the 
four-Fermi operators, denoted by ${\cal O}_9$ and ${\cal O}_{10}$:
\bea
{\cal O}_9    & \sim & (\bar{s}_L\gamma_{\mu} b_L)
                \sum_\ell(\bar{\ell}\gamma^{\mu}\ell) \, , \nonumber\\
    O_{10} & \sim & (\bar{s}_L\gamma_{\mu} b_L)
                \sum_\ell(\bar{\ell}\gamma^{\mu} \gamma_{5} \ell) \, ,
\eea
weighted, respectively, by the corresponding Wilson coefficients
$C_9(\mu)$ and $C_{10}$. In most supersymmetric theories, the
SM basis is sufficient to describe the generic transitions $b \to s
\g$ and $b \to s \ell^+ \ell^-$, and we shall confine ourselves to
discussing some possible BSM effects in this context.

 Experimentally, the goal
is to precisely measure a number of differential distributions, of which 
the
dilepton invariant mass spectrum and the forward-backward asymmetry of the
charged leptons in the dilepton rest frame are the best studied
theoretically. BSM physics could manifest itself through 
additional contributions in the Wilson coefficients, shifting their values
from the ones in the SM. This will lead to possible distortions in the
two mentioned decay distributions, as well as some others not discussed 
for lack of space. From a theoretical point of view,
inclusive decays $B \to X_s \ell^+ \ell^-$ are more robust, in particular
in the dilepton mass region below the $J/\psi$ resonance, as
the explicit ${\cal O}(\alpha_s)$ improvements in the dilepton
invariant mass distributions are now available in this
region. Furthermore, the long-distance contributions, implemented through
the matrix elements of the operators ${\cal O}_1$ and ${\cal O}_2$, can 
be brought under control by a judicious choice of the experimental cuts, 
or estimated theoretically~\cite{Ali:1996bm,Buchalla:1998ky}.

The dilepton invariant mass distribution
for the inclusive decay $B \to X_s \ell^+ \ell^-$ can be written as
\begin{eqnarray}
    &&\frac{d\Gamma(b\to X_s \ell^+\ell^-)}{d\hat s}=
    \left(\frac{\alpha_{em}}{4\pi}\right)^2
    \frac{G_F^2 m_{b,pole}^5\left|V_{ts}^*V_{tb}^{}\right|^2}
    {48\pi^3}(1-\hat s)^2 \times \label{rarewidth} \\
    &&\left ( \left (1+2\hat s\right)
    \left (\left |\widetilde C_9^{\eff}\right |^2+
    \left |\widetilde C_{10}^{\eff}\right |^2 \right )   
    + 4\left(1+2/\hat s\right)\left
    |\widetilde C_7^{\eff}\right |^2+
    12 \mbox{Re}\left (\widetilde C_7^{\eff}
    \widetilde C_9^{\eff*}\right ) \right ) \, \nn.
\end{eqnarray}  
The effective Wilson coefficients  $\tilde{C}_7^{\eff}$,
$\tilde{C}_9^{\eff}$ and $\tilde{C}_{10}^{\eff}$, including explicit
$O(\alphas)$ corrections, have been calculated~\cite{Ali:2001jg,BMU,AAGW}.

\begin{figure}[H]
\begin{center}
\epsfig{file=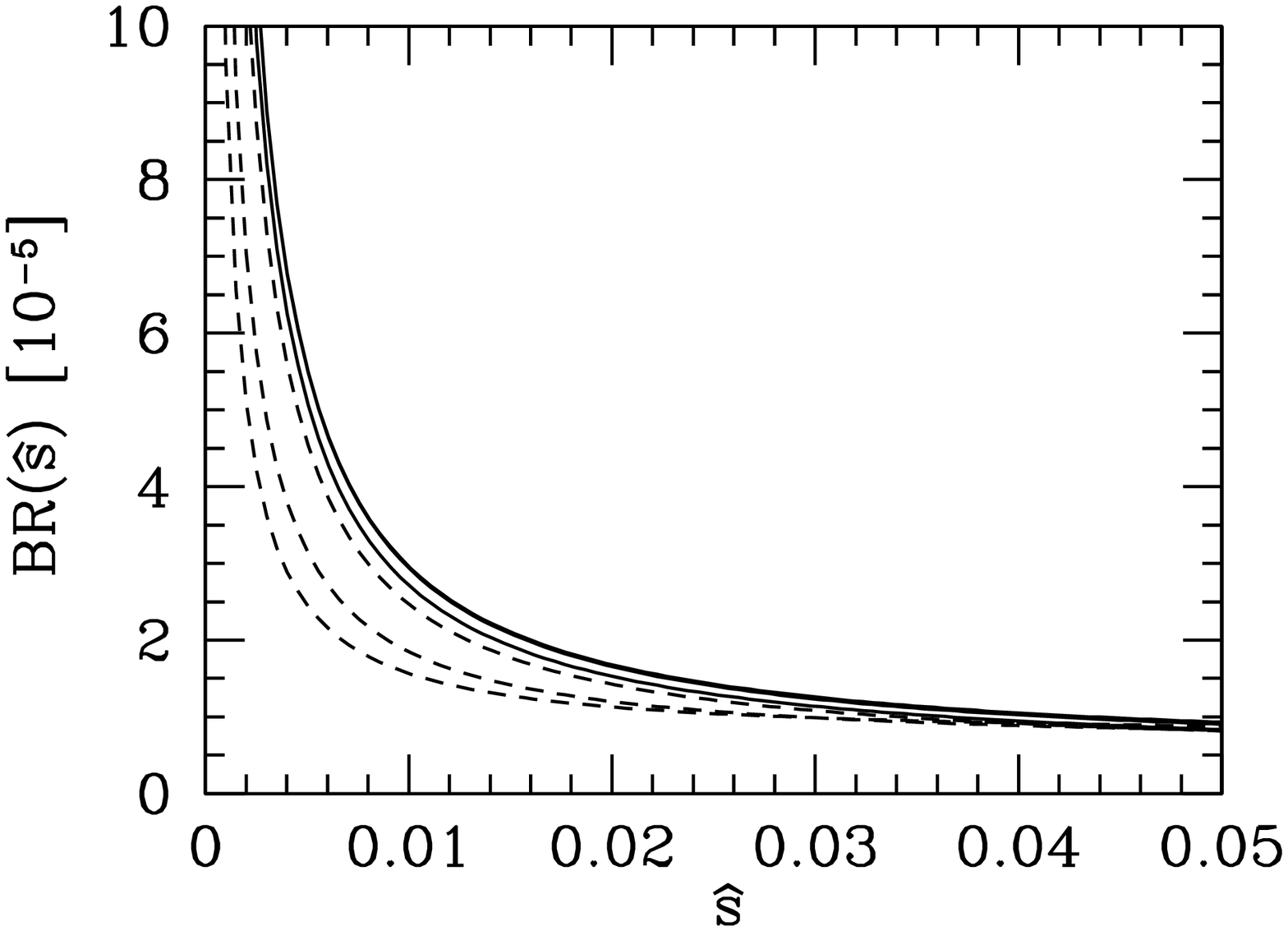,width=0.48\linewidth}
\hspace*{.2cm}
\epsfig{file=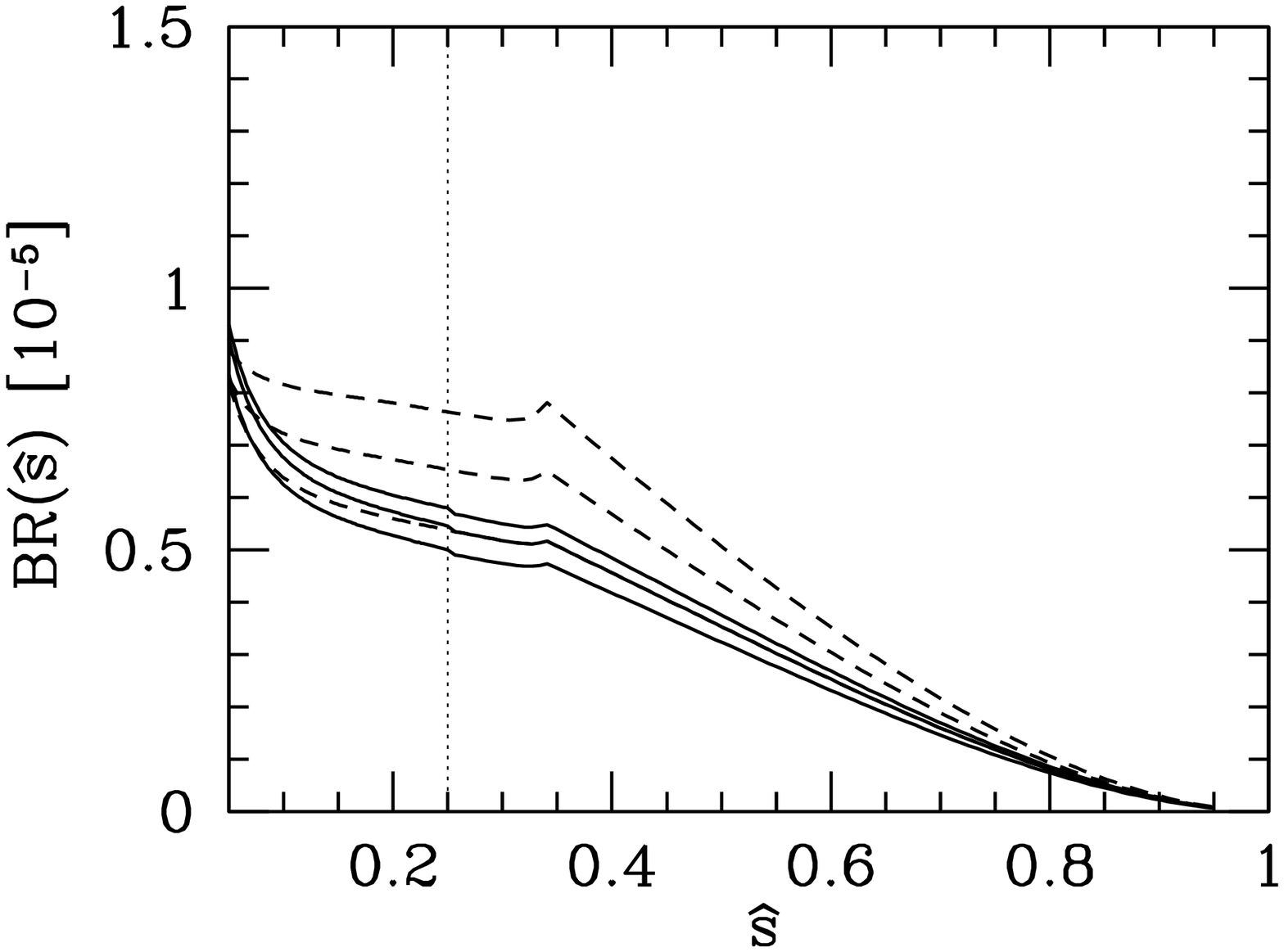,width=0.48\linewidth}
\caption{\it Partial (dashed lines) vs full (solid lines) NNLO 
computation of the branching ratio $B\to X_s e^+ e^-$. In the
left plot ($\hat s \in [0,0.05]$) the lowest curves are for
$\m=10\;\gev$ and the uppermost ones for $\mu=2.5\;\gev$. In the right
plot the $\m$ dependence is reversed: the uppermost curves correspond
to $\m=10\;\gev$ and the lowest ones to $\mu=2.5\;\gev$. The right-hand
plot also holds for the decay $B \to X_s \mu^+ \mu^-$. 
(From Ref.~15.)}
\label{fig:nnlo}
\end{center}
\end{figure}

The dilepton invariant mass distribution for the process $B \to X_s
e^+ e^-$ calculated in  NNLO is shown in \fig{fig:nnlo} for
the three choices of the scale $\mu=2.5~\gev$, $\mu=5~\gev$ and
$\mu=~10~\gev$ (solid curves). In this figure, the left-hand plot
shows the distribution in the very low invariant mass region ($\hat s
\in [0,0.05]$, with $0$ to be understood as the kinematic threshold $s =4
m_e^2 \simeq 10^{-6}\; \gev^2$, yielding $\hat{s}=3.7 \times
10^{-8}$), and the right-hand plot shows the dilepton
spectrum in the region beyond $\hat {s} > 0.05$, and hence this also
holds for the decay $B \to X_s \mu^+ \mu^-$. It should be stressed 
that a genuine NNLO calculation only exists for values
of $\hat{s}$ below 0.25, which is indicated in the right-hand plot
by the vertical dotted line. For higher values of $\hat{s}$, an
estimate of the NNLO result is obtained by an extrapolation procedure
discussed in detail elsewhere \cite{Ali:2001jg}. 
The so-called partial NNLO dilepton spectrum
is also shown in each of these cases
for the same three choices of the scale $\mu$ (dashed curves).
Note that the NNLO
dilepton invariant mass spectrum in the right-hand plot ($\hat {s} >
0.05$) lies below its partial NNLO counterpart, and hence the partial
branching ratios for both the $B \to X_s e^+ e^-$ and $B \to X_s \mu^+
\mu^-$ decays are reduced in the full NNLO accuracy.

This framework has recently been used to calculate the branching ratio
for $B \to X_s \ell^+ \ell^-$ in the SM, yielding~\cite{Ali:2001jg}
\bea
\branch (B\to X_s e^+ e^-) &=& (6.9 \pm 1.0) \times 10^{-6} \;\;\;
        (\d \branch_{X_see} = \pm 15 \%) \; , \\
\branch (B\to X_s \m^+ \m^-) &=& (4.2 \pm 0.7) \times 10^{-6} \;\;\;
        (\d \branch_{X_s\m\m} = \pm 17 \%) \; .
\eea
The theoretical errors shown are obtained by estimating the errors from
the individual input parameters, and the details can be seen in the 
original work. Current
experimental data sees no signal for these decays, 
yielding the following upper limits\cite{bellebsll}
\bea
\branch (B\to X_s \mu^+ \mu^-) &\leq & 19.1 \times  10^{-6} \; {\rm at}\;
 \cl{90} \;   \label{bsmmexp} ,\\
\branch (B\to X_s e^+ e^-) &\leq& 10.1 \times  10^{-6} \; {\rm
at}\; \cl{90}
 \;  . \label{bseeexp}
\eea 
Thus, the current experimental sensitivity is typically a factor 3 away 
from the SM-estimates. 
We now
turn to the modifications of the effective Wilson coefficients
$\widetilde C_7^{\eff}$, $\widetilde C_9^{\eff}$ and $\widetilde
C_{10}^{\eff}$ in the presence of new physics which modifies
the Wilson coefficients $C_7$, $C_8$, $C_9$ and
$C_{10}$ at the matching scale $\mu_W$. For lack of complete NLO 
calculations, we assume that only the lowest non-trivial order of these   
Wilson coefficients get modified by new physics, which  means that
$C_7^{(1)}(\mu_W)$, $C_8^{(1)}(\mu_W)$, $C_9^{(1)}(\mu_W)$,
$C_{10}^{(1)}(\mu_W)$ get only indirectly modified. The shifts of the 
Wilson
coefficients at the scale $\mu_W$ can be written as ($i=7$,...,$10$):
\begin{equation}
C_i(\mu_W) \longrightarrow C_i(\mu_W) + \frac{\alpha_s}{4\pi} \,  
C_i^{NP}(\mu_W) \, .
\end{equation}
These shift at the matching scale are translated through the RGE step
into modifications of the coefficients $C_i (\m_b)$ at the low scale
$\mu_b$. The bounds implied by the experimental results given in
Eq.~(\ref{bseeexp}) (being the more stringent of the two limits) have 
been computed in the $[C_9^{NP} (\mu_W),C_{10}^{NP}]$
plane~\cite{Ali:2001jg}. We shall show the cumulative bounds resulting 
from the combined analysis of all the inclusive and exclusive semileptonic 
decay in the next section.
\section{Exclusive Decays $B \to K^* \gamma$ and $B \to K^* \ell^+ 
\ell^-$ in the SM\\ and Supersymmetry}
\label{sec:exclusives}
 Concentrating first on the transitions $B \to
K^* \gamma^{(*)}$, with a real or virtual photon, the general
decomposition of the matrix elements
on all possible Lorentz structures present in the effective
Hamiltonian admits seven form factors, which for the dilepton final state
are functions of the momentum squared~$q^2$ transferred from the heavy
meson to the light one. When the energy of the final light meson~$E$
is large (the large recoil limit), one can expand the interaction of the
energetic quark in the meson with the soft gluons in terms of  
$\Lambda_{\rm QCD}/E$. Using HQET for the interaction of the
heavy $b$-quark with the gluons, one can derive non-trivial relations
between the soft contributions to the form factors \cite{Charles:1999dr}.
The resulting theory (LEET) reduces the number of
independent form factors from seven in the $B \to K^* \gamma^*$ 
transitions to two in this limit. The relations among the form factors in 
the symmetry limit are broken by perturbative  QCD radiative corrections
arising from the vertex renormalization and the hard spectator
interactions~\cite{Beneke:2001wa}.
To incorporate  both types of QCD corrections, a 
factorization formula for the heavy-light form factors at large   
recoil is useful~\cite{Beneke:2001wa}:
\begin{equation}
f_k (q^2) = C_{\perp k} \xi_\perp +  C_{\| k} \xi_\| +
\Phi_B \otimes T_k \otimes \Phi_\rho ,
\label{eq:fact-formula}
\end{equation}
where $f_k (q^2)$ is any of the seven independent form factors in
the $B \to K^*$ transitions at hand; $\xi_\perp$ and~$\xi_\|$ are
the two independent form factors remaining in the LEET-symmetry
limit; $T_k$ is a hard-scattering kernel calculated in
$O (\alpha_s)$;
$\Phi_B$ and~$\Phi_\rho$ are the light-cone distribution
amplitudes of the~$B$- and~$\rho$-meson convoluted with~$T_k$;
$C_k = 1 + O (\alpha_s)$ are the hard vertex renormalization
coefficients. An $O(\alpha_s)$ proof of the validity of
Eq.~(\ref{eq:fact-formula}) for radiative decays has, in the meanwhile,  
been provided by several 
groups~\cite{Bosch:2001gv,Ali:2001ez,Beneke:2001at}, yielding
\begin{eqnarray}
{\cal B}_{\rm th} (B \to K^{*} \gamma) & = &
\tau_B \, \Gamma_{\rm th} (B \to K^* \gamma)
\label{eq:DW(B-Kgam)} \\
& = &
\tau_B \,\frac{G_F^2 \alpha |V_{tb} V_{ts}^*|^2}{32 \pi^4} \,
m_{b, {\rm pole}}^2 \, M^3 \, \left [ \xi_\perp^{(K^*)} \right ]^2
\left ( 1 - \frac{m_{K^*}^2}{M^2} \right )^3 \nonumber\\
& \times & \left | C^{(0){\rm eff}}_7 +  A^{(1)}(\mu) \right |^2 ,
\nonumber
\end{eqnarray}
where~$\alpha = \alpha(0)=1/137$ is the fine-structure constant,
$M$~and $m_{K^*}$ are the $B$- and $K^*$-meson masses,
and~$\tau_B$ is the lifetime of the~$B^0$- or $B^+$-meson.
The function~$ A^{(1)}$ in Eq.~(\ref{eq:DW(B-Kgam)})
lumps all three explicit $O(\alpha_s)$ contributions from
the Wilson coefficient~$C_7^{\rm eff}$, $b \to s \gamma$ vertex, 
and the hard-spectator corrections to the $B \to K^* \gamma$ amplitude.
NLO corrections yield a typical "K-factor" of 
1.6 ,yielding~\cite{Ali:2001ez} 

\bea
{\cal B}_{\rm th} (B \to K^* \gamma) & \simeq &
(7.2 \pm 1.1)\times 10^{-5} \,
\left ( \frac{\tau_B}{1.6~{\rm ps}} \right )
\left ( \frac{m_{b,{\rm pole}}}{4.65~{\rm GeV}} \right )^2
\left ( \frac{\xi_\perp^{(K^*)}}{0.35} \right )^2 \nonumber\\  
& = & (7.2 \pm 2.7) \times 10^{-5}, 
\label{eq:Br-Ksgam}
\eea
where the enlarged error in the second equation reflects the assumed error
in the nonperturbative quantity, $\xi_\perp^{(K^*)} (0) = 0.35 \pm 0.07$.
The estimates presented in Refs.~\cite{Beneke:2001at,Bosch:2001gv}
are similar.The LEET-based estimates are  larger than
the experimental branching ratio for $B \to K^* \gamma$:
\begin{eqnarray}
{\cal B} (B^\pm \to K^{* \pm} \gamma) =
(3.82 \pm 0.47) \times 10^{-5},
\label{eq:Br-Ks-exp} \\
{\cal B} (B^0 (\bar B^0) \to K^{* 0}(\bar K^{*0}) \gamma) =
(4.44 \pm 0.35) \times 10^{-5}~,
\nonumber
\end{eqnarray}
though the
attendant theoretical error, estimated as~$\pm 40\%$, does not allow
to draw a completely quantitative conclusion.

The price of agreement between the LEET approach and data can be
specified in terms of the LEET-form factor $\xi_\perp^{(K^*)} (0)$.
To that end, it is advantageous to
calculate the following ratio of the exclusive to inclusive branching 
ratios: 
\begin{equation}
R_{\rm exp} (K^* \gamma/X_s \gamma) \equiv
\frac{\bar {\cal B}_{\rm exp} (B \to K^* \gamma)}
     {\bar {\cal B}_{\rm exp} (B \to X_s \gamma)} = 0.13 \pm 0.02 ,
\label{eq:Ks/Xs-exp}
\end{equation}  
where the current experimental value of this ratio is also given,
averaging over the charged and neutral $B$-decays.
At NLO, the ratio $R(K^*\gamma/X_s \gamma)$ yields~\cite{Ali:2001ez}
\begin{equation}
\bar \xi_\perp^{(K^*)} (0) = 0.25 \pm 0.04 , \qquad
\left [ \bar T_1^{(K^*)} (0, \bar m_b) = 0.27 \pm 0.04 \right ]~,
\label{eq:xi-Ks-average}
\end{equation}
where $\bar T_1^{(K^*)} (0, \bar m_b)$ is the form factor in full QCD,
determined using the quark masses in the $\overline{MS}$-scheme.
This values is significantly smaller than
the corresponding predictions from the QCD sum rules 
analysis~\cite{Ali:2000mm,Ball:1998kk}
$T_1^{(K^*)} (0) = 0.38 \pm 0.06$, and from the lattice 
simulations~\cite{DelDebbio:1998kr}
$T_1^{(K^*)} (0) = 0.32^{+0.04}_{-0.02}$.
The reason for this mismatch is not obvious and this point deserves
further theoretical study.

In view of this, one can not insist that the absolute rates in exclusive 
decays can be calculated reliably in the LEET-approach. 
It should, however, be emphasized that any measurable CP asymmetry in the 
exclusive ($B \to K^* \gamma)$ or inclusive ( $B \to X_s \gamma$) 
decay will be a sure sign of BSM physics, as the SM CP asymmetry in either of 
these modes~\cite{Soares:1991te} is not expected
to exceed $\frac{1}{2}\%$. Present experimental bounds on the CP 
asymmetry
are \cite{Coan:2001cp}${\cal A}_{\rm CP}(B \to X_s \gamma) =(-0.079 \pm 
0.018 \pm 0.022)$, and~\cite{Ryd:2001cp} ${\cal A}_{\rm CP}(B \to K^{*0} 
\gamma) =(-0.035 \pm 0.094 \pm 0.012)$, obtained in the $K^+\pi^-$ mode.
The first error in both cases is statistical and the second systematic.
They still allow a lot of room for the BSM physics; however, not in the
MFV and Extended-MFV models, discussed earlier. 

\subsection{$B \to K^{*} \ell^+ \ell^-$ Decays}
\label{ssec:B-kstar-ll}
The NNLO corrections for $B \to X_s \ell^+ \ell^-$ calculated by Bobeth et 
al.~\cite{BMU}
and by Asatrian et al.~\cite{AAGW} for the short-distance
contribution have been recently harnessed~\cite{Ali:2001jg}
to study the exclusive decays $B\to K^{(*)} \ell^+ \ell^-$.
This input is then combined with the 
form factors calculated with the help of the QCD sum rules~\cite{Ali:2000mm},
ignoring the so-called hard spectator corrections, calculated in the 
decays $B \to K^* \ell^+ \ell^-$ \cite{Beneke:2001at}.
The rationale of this is the following:  Beneke et 
al.~\cite{Beneke:2001at} have shown that the dilepton invariant mass 
distribution in the low invariant mass  
region is rather stable against the explicit $O(\alpha_s)$ corrections, 
and the theoretical uncertainties are dominated by the form factors
and other non-perturbative parameters specific to the large-energy
factorization approach. As already discussed, current
data on $B \to K^* \gamma$ decay yields typically a  range
$T_1(0)=0.27 \pm 0.04$. To accommodate this, a value 
$T_1(0)=0.33 \pm 0.05$, corresponding to the lower set of values in the
QCD sum rules have been used in the NNLO analysis~\cite{Ali:2001jg},
yielding
\bea
\branch (B\to K \ell^+ \ell^-) &=& (0.35 \pm 0.12) \times 10^{-6} \;\;\;
        (\d \branch_{K\ell\ell} = \pm 34 \%) \; , \\
\branch (B\to K^* e^+ e^-) &=& (1.58 \pm 0.49) \times 10^{-6} \;\;\;
        (\d \branch_{K^*ee} = \pm 31 \%) \; , \\
\branch (B\to K^* \m^+ \m^-) &=& (1.19 \pm 0.39) \times 10^{-6} \;\;\;
        (\d \branch_{K^*\m\m} = \pm 33 \%) \; .
\eea
These estimates are lower than the NLO estimates~\cite{Ali:2000mm} due to 
two reasons: the explicit $O(\alpha_s)$ corrections lower the decay rates 
and the central values
of the input form factors are also reduced so as to accommodate the $B \to 
K^* \gamma$ branching ratios in the same accuracy in $O(\alpha_s)$.
They have to be confronted with the BELLE data~\cite{Abe:2001dh} 
summarized below:
\bea
\branch (B\to K \ell^+ \ell^-) &=& (0.75^{+0.25}_{-0.21}\pm 0.09)\times
10^{-6} \; , \label{bkllexp} \\
\branch (B\to K^* \mu^+ \mu^-) &\leq& 3.0 \times  10^{-6} \; {\rm at}\;
 \cl{90}  \; \label{bksmmexp} , \\
\branch (B\to K^* e^+ e^-) &\leq& 5.1 \times  10^{-6} \; {\rm at}\;
 \cl{90} \;  \label{bkseeexp} .
\eea
Very recently, upper limits on these decays have been set by the BABAR 
collaboration, which at $90\%$ C.L. are posted as~\cite{Aubert:2002aw}
\bea
\branch (B\to K \ell^+ \ell^-) &\leq& 0.50 \times 10^{-6} \; , 
\label{bkllexp-babar} \\
\branch (B\to K^* \ell^+ \ell^-) &\leq& 2.9 \times 10^{-6} \; ,
\label{bksllexp-babar}    
\eea
where a ratio $\branch (B\to K^* e^+ e^-)/\branch (B\to K^* \mu^+ 
\mu^-)=1.2$, following from the NLO QCD-SR estimate~\cite{Ali:2000mm} has 
been used to combine the $K^*e^+e^-$ and $K^* \mu^+ \mu^-$ modes. 
As opposed to the BELLE collaboration, reporting a statistically 
significant signal in the $B \to K \ell^+\ell^-$ modes, BABAR data has no 
signal in this mode. However, the BABAR upper limit is not inconsistent 
with the 
BELLE measurement fluctuated down by slightly over a standard deviation.
  
\begin{figure}[t]
\begin{center}
\epsfig{file=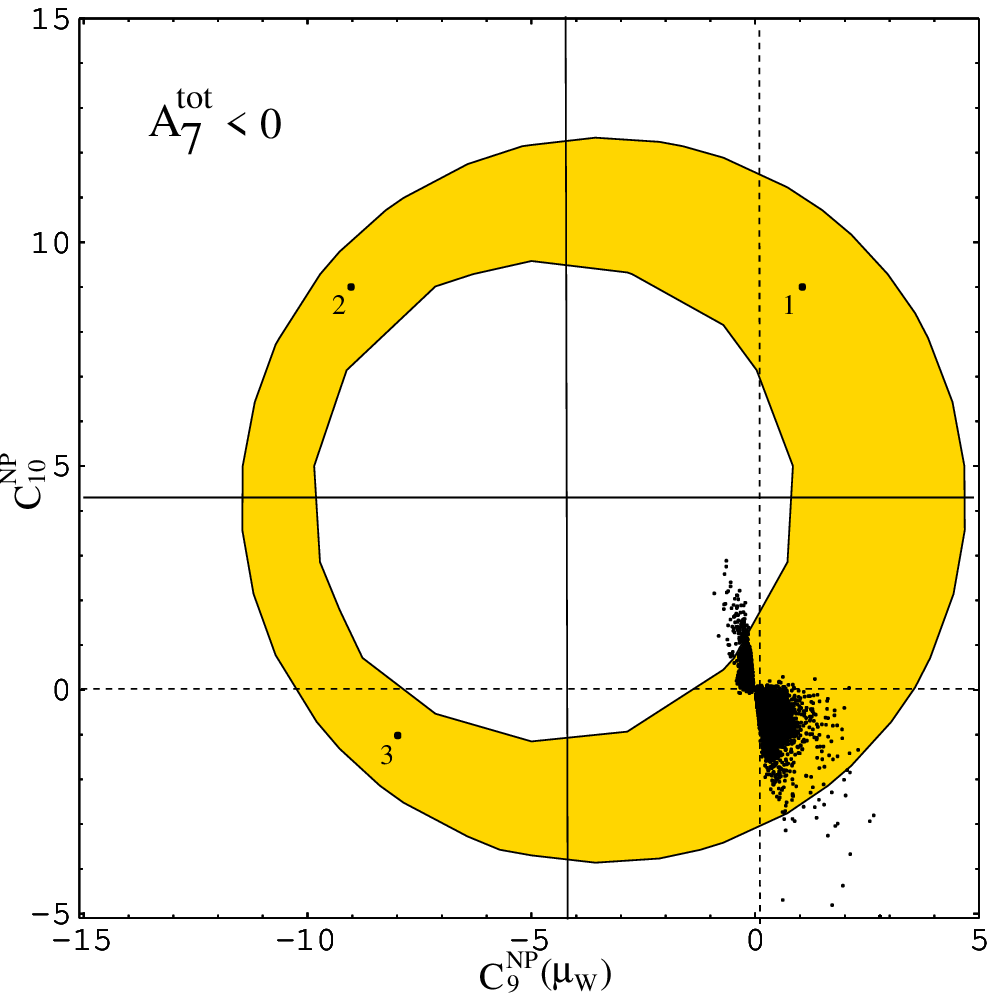,width=0.4\linewidth}
\hskip 0.5cm
\epsfig{file=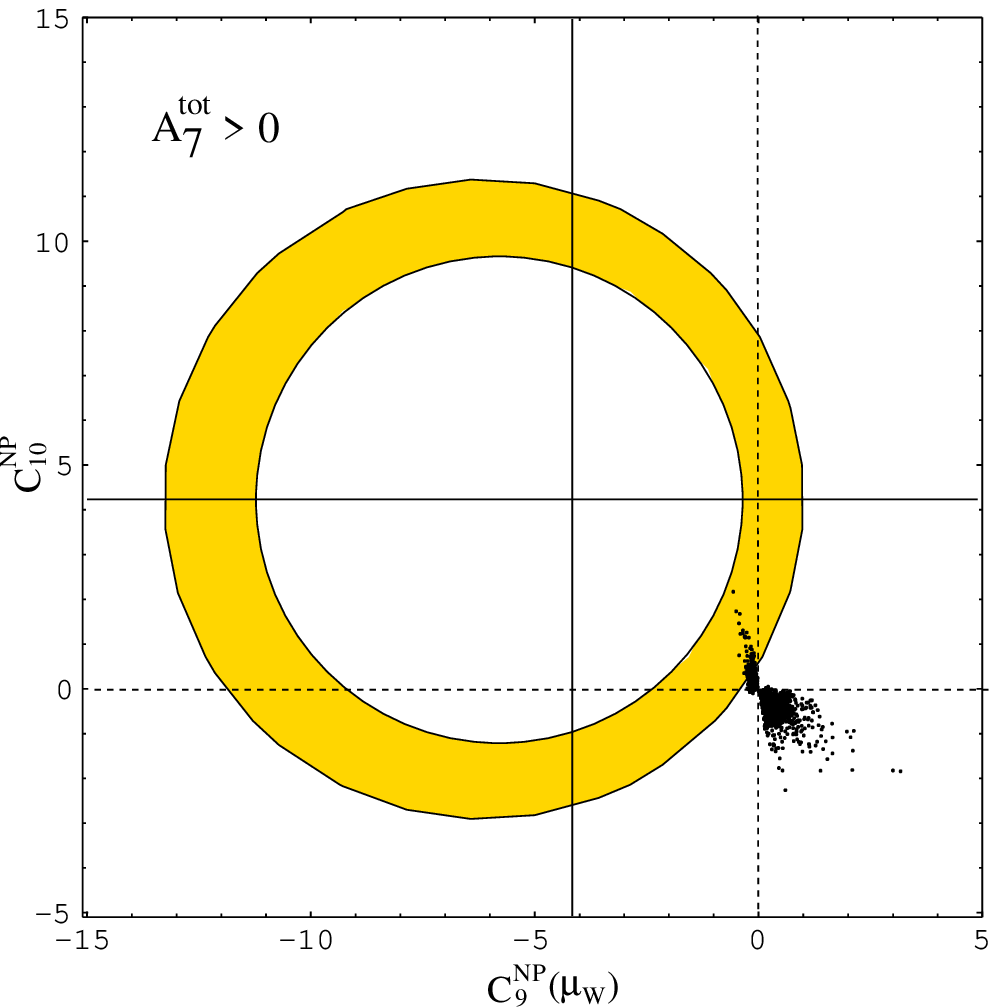,width=0.4\linewidth}
\caption{\it {\bf NNLO Case.} Superposition of all the
constraints. The plots correspond to the $C_7^{\rm tot}(2.5 \;
\gev)<0$ and $C_7^{\rm tot}(2.5 \; \gev) >0$ case, respectively. The
points are obtained by means of a scanning over the EMFV parameter
space and requiring the experimental bound from $B\to X_s \g$ to be
satisfied. (From Ref.~15.)}
\label{fig:total}
\end{center}
\end{figure}

\begin{figure}[H]
\begin{center}
\epsfig{file=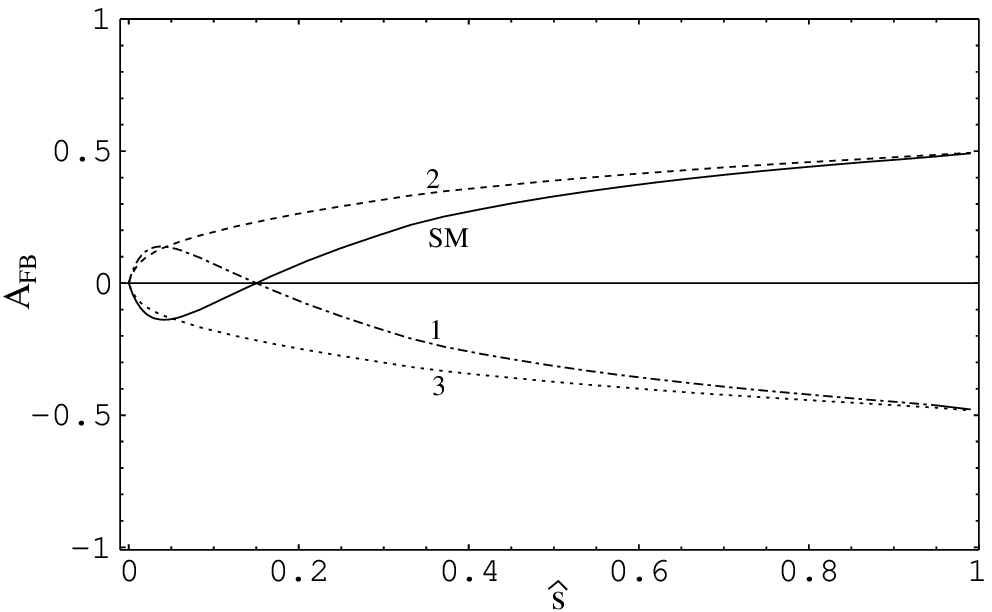,width=0.7\linewidth}
\caption{\it Differential Forward--Backward asymmetry for the decay
$B\to X_s \ell^+ \ell^-$. The four curves correspond to the points
indicated in \fig{fig:total}. (From Ref.~15.)}
\label{fig:afb}
\end{center}
\end{figure}

To quantify the agreement between the SM and current data, 
we show the bounds implied by the experimental results given above 
in the $[C_9^{NP} (\mu_W),C_{10}^{NP}]$ plane. 
 In \fig{fig:total}, the bounds
from the inclusive radiative decays and inclusive and exclusive 
semileptonic decays have been combined in a
single plot. Note that the overall allowed region is driven by the 
constraints emanating from the decays $B\to X_s e^+ e^-$ and $B\to K 
\ell^+ \ell^-$. In showing the constraints  from
$B \to K \ell^+ \ell^-$, we have used the BELLE measurement to get the
following bounds:
\be
0.37 \; \times 10^{-6} \leq \branch (B\to K
\ell^+ \ell^-) \leq 1.2 \; \times 10^{-6} \; {\rm at} \; \cl{90} \; ,    
\label{BKLLBR}
\ee
resulting in carving out an inner region in the $(C_9^{\rm 
NP}(\m_W),C_{10}^{\rm NP})$ plane. The two plots shown in these figures
correspond respectively to the $C_7^{\rm tot}$-negative and $C_7^{\rm
tot}$-positive solutions discussed earlier. In \fig{fig:total} four 
regions are identified which are allowed  by the
constraints on the branching ratios that present very different
forward--backward asymmetries. In \fig{fig:afb} we show the shape of
the FB asymmetry spectrum for the SM and other three sample
points. The distinctive features are the presence or not of a zero and
global sign of the asymmetry. A rough indication of the FB asymmetry
behavior is thus enough to rule out a large part of the parameter
space that the current branching ratios can not explore. 

In order to explore the region in the $[C_9^{NP},C_{10}^{NP}]$ plane
(where $C_{9,10}^{NP}$ are the sum of MFV and MI contributions)
 that is
accessible to these models, a high statistic scanning
over the EMFV parameter space has been recently 
performed~\cite{Ali:2001jg}
requiring 
each point to survive the
constraints coming from the sparticle masses lower bounds and $b\to s
\g$. The surviving points are shown in \fig{fig:total} together with
the model independent constraints. Note that the region spanned by
these points has been drastically reduced by the presence of the $b\to
s \g$ constraint.
\section{Summary}

We summarize the main points of this talk.

\bit
\item SM is in comfortable agreement with the measurements of the 
CP-asymmetry $a_{J/\psi}K_s$, yielding $\sin 2 \beta$. However, current 
data also allows a BSM phase, with a typical range $-3^\circ < \theta_d < 8^\circ$.
\item SM is also in comfortable agreement with data on $B \to X_s \gamma$.  
Due to the inherent two-fold ambiguity on the sign of the effective Wilson
coefficient, both $C_7^{\rm tot}>0$ and $C_7^{\rm tot}<0$ solutions are
allowed in supersymmetry involving different regions of the parameter
space.
\item Theoretical precision in exclusive decays
is at present compromised by the imprecise knowledge of form factors and some
other non-perturbative quantities. Ratios of the branching ratios, and some
asymmetries (due to isospin-violations or CP-violation in $B \to \rho
\gamma$ and $B \to K^* \gamma$) are more reliably calculable in this
framework, and can be used to search for BSM physics. A quantitative test of
the SM in these decays will be undertaken
in inclusive decays $B \to X_s \ell^+ \ell^-$.
\item Despite theoretical uncertainties, the experimental sensitivity on 
rare semileptonic $B$ decays is 
already strong enough to provide non trivial bounds on the SUSY   
parameter space. Indeed, for the $C_7^{\rm tot}>0$ case, the larger  
portion of the SUSY allowed points is already ruled out.
\item SUSY models can account only for a
small part of the region allowed by the model independent
analysis of current data. In the numerical analysis
discussed here \cite{Ali:2001jg}, integrated branching 
ratios have been used to put
constraints on the effective coefficients. They allow a multitude
of solutions in the effective Wilson parameter space and can be disentangled
from each other only with the help of both the dilepton mass spectrum and the
forward-backward asymmetryin semileptonic rare $B$ decays.
Only such measurements would allow us to determine the exact values and
signs of the Wilson coefficients $C_7$, $C_9$ and $C_{10}$, also limiting
$C_8$ and decipher the physics behind flavour transitions.
\eit

\section*{Acknowledgments}
I would like to thank Cai-Dian L\"u and Yue-Liang Wu for their 
warm hospitality during my stay in Beijing and at the ICFP 2001 
conference.


\section*{References}
\begin{thebibliography}{99}
\bibitem{Aubert:2001cp}
B.~Aubert {\em et~al.} [BABAR Collaboration],
Phys.\ Rev.\ Lett.\  {\bf 87} (2001) 091801
[hep-ex/0107013].

\bibitem{Abashian:2001cp}
K.~Abe {\em et~al.} [BELLE Collaboration],
Phys.\ Rev.\ Lett.\  {\bf 87} (2001) 091802
[hep-ex/0107061].

\bibitem{Alam:1995aw}
M.~S.~Alam {\it et al.}  [CLEO Collaboration],
Phys.\ Rev.\ Lett.\  {\bf 74} (1995) 2885.

\bibitem{cleobsg}   
S. Chen {\it et al.} [CLEO Collaboration],
Phys.\ Rev.\ Lett.\  {\bf 87} (2001) 251807 [hep-ex/0108032].

\bibitem{alephbsg}
R.~Barate {\em et~al.} [ALEPH Collaboration],
\newblock Phys. Lett. {\bf B429}, 169 (1998).

\bibitem{bellebsg}
K.~Abe {\it et al.} [BELLE Collaboration],
Phys.\ Lett.\ B {\bf 511} (2001) 151
[hep-ex/0103042].

\bibitem{Chetyrkin:1997vx}  
K.~Chetyrkin, M.~Misiak and M.~M\"unz,
Phys.\ Lett.\ B {\bf 400}, 206 (1997)
[Erratum-ibid.\ B {\bf 425}, 414 (1997)]
[hep-ph/9612313].

\bibitem{Kagan:1999ym} 
A.~L.~Kagan and M.~Neubert,
Eur.\ Phys.\ J.\ C {\bf 7}, 5 (1999)
[hep-ph/9805303].

\bibitem{Gambino:2001ew}
P.~Gambino and M.~Misiak,
Nucl.\ Phys.\ B {\bf 611}, 338 (2001)
[hep-ph/0104034].


\bibitem{bellebsll}    
K.~Abe {\it et al.}  [Belle Collaboration],
BELLE-CONF-0110 [hep-ex/0107072].

\bibitem{Abe:2001dh} 
K.~Abe {\it et al.}  [BELLE Collaboration],
Phys.\ Rev.\ Lett.\  {\bf 88} (2002) 021801
[hep-ex/0109026].

\bibitem{Aubert:2002aw}
B.~Aubert et al.~[BABAR Collaboration],
Report BABAR-PUB-01/19; SLAC-PUB-9102 [hep-ex/0201008].

\bibitem{Ali:1996bm}
A.~Ali, G.~Hiller, L.~T.~Handoko and T.~Morozumi,
Phys.\ Rev.\ D {\bf 55}, 4105 (1997)
[hep-ph/9609449].

\bibitem{Ali:2000mm}
A.~Ali, P.~Ball, L.~T.~Handoko and G.~Hiller,
Phys.\ Rev.\ D {\bf 61}, 074024 (2000)
[hep-ph/9910221].

\bibitem{Ali:2001jg}
A.~Ali, E.~Lunghi, C.~Greub and G.~Hiller,
Report DESY 01-217 [hep-ph/0112300].



\bibitem{BMU}
C.~Bobeth, M.~Misiak and J.~Urban,
Nucl.\ Phys.\ B {\bf 574} (2000) 291
[hep-ph/9910220].

\bibitem{AAGW}
H.~H.~Asatrian, H.~M.~Asatrian, C.~Greub and M.~Walker,
Phys.\ Lett.\ B {\bf 507} (2001) 162
[hep-ph/0103087]; \\
H.~H.~Asatryan, H.~M.~Asatrian, C.~Greub and M.~Walker,
[hep-ph/0109140].

\bibitem{ali1}
See, for example, A.~Ali and D.~London,
\newblock Eur. Phys. J. {\bf C18}, 665 (2001), [hep-ph/0012155].

\bibitem{giudice}
M.~Ciuchini, G.~Degrassi, P.~Gambino, and G.~F. Giudice,
\newblock Nucl. Phys. {\bf B534}, 3 (1998), [hep-ph/9806308].

\bibitem{ali2}
A.~Ali and D.~London,
\newblock Eur. Phys. J. {\bf C9}, 687 (1999), [hep-ph/9903535].
\newblock Phys. Rept. {\bf 320}, 79 (1999), [hep-ph/9907243].


\bibitem{nelson}
A.~G. Cohen, D.~B. Kaplan, F.~Lepeintre, and A.~E. Nelson,
\newblock Phys. Rev. Lett. {\bf 78}, 2300 (1997), [hep-ph/9610252].

\bibitem{wolf-silva}
J.~P. Silva and L.~Wolfenstein,
\newblock Phys. Rev. {\bf D55}, 5331 (1997), [hep-ph/9610208].

\bibitem{mia}
L.~J. Hall, V.~A. Kostelecky, and S.~Raby,
\newblock Nucl. Phys. {\bf B267}, 415 (1986).

\bibitem{mia2}
A.~J. Buras, A.~Romanino, and L.~Silvestrini,
\newblock Nucl. Phys. {\bf B520}, 3 (1998), [hep-ph/9712398].

\bibitem{Ali:2001ej}
A.~Ali and E.~Lunghi,
Eur.\ Phys.\ J.\ C {\bf 21}, 683 (2001)
[hep-ph/0105200].

\bibitem{Ali:2000zu}
A.~Ali, L.~T.~Handoko and D.~London,
Phys.\ Rev.\ D {\bf 63}, 014014 (2000)
[hep-ph/0006175].

\bibitem{Bosch:2001gv}
S.~W.~Bosch and G.~Buchalla,
Nucl.\ Phys.\ B {\bf 621}, 459 (2002)
[hep-ph/0106081]. 


\bibitem{Ali:2001ez} 
A.~Ali and A.~Y.~Parkhomenko,
Report DESY 01-068 [hep-ph/0105302].

\bibitem{Charles:1999dr}
J.~Charles, A.~Le Yaouanc, L.~Oliver, O.~Pene and J.~C.~Raynal,
Phys.\ Rev.\ D {\bf 60}, 014001 (1999)
[hep-ph/9812358].

\bibitem{Hewett:1996ct}
J.~L.~Hewett and J.~D.~Wells,
Phys.\ Rev.\ D {\bf 55}, 5549 (1997)
[hep-ph/9610323].

\bibitem{LMSS}
E.~Lunghi, A.~Masiero, I.~Scimemi and L.~Silvestrini,
Nucl.\ Phys.\ B {\bf 568} (2000) 120
[hep-ph/9906286].

\bibitem{Buchalla:1998ky}
G.~Buchalla, G.~Isidori and S.~J.~Rey,
Nucl.\ Phys.\ B {\bf 511}, 594 (1998)
[hep-ph/9705253].

\bibitem{Beneke:2001wa}
M.~Beneke and T.~Feldmann,
Nucl.\ Phys.\ B {\bf 592}, 3 (2001)
[hep-ph/0008255].


\bibitem{Beneke:2001at}
M.~Beneke, T.~Feldmann and D.~Seidel,
Nucl.\ Phys.\ B {\bf 612}, 25 (2001)
[hep-ph/0106067].

\bibitem{Ball:1998kk}
P.~Ball and V.~M.~Braun,
Phys.\ Rev.\ D {\bf 58}, 094016 (1998)
[hep-ph/9805422].

\bibitem{DelDebbio:1998kr}
L.~Del Debbio, J.~M.~Flynn, L.~Lellouch and J.~Nieves
[UKQCD Collaboration],
Phys.\ Lett.\ B {\bf 416}, 392 (1998)
[hep-lat/9708008].


\bibitem{Soares:1991te}
J.~M.~Soares,
Nucl.\ Phys.\ B {\bf 367}, 575 (1991).

\bibitem{Coan:2001cp}
T.~Coan et al. [CLEO Collaboration],
Phys.\ Rev.\ Lett.\  {\bf 86}, 5661 (2001)
[hep-ex/0010075].

\bibitem{Ryd:2001cp}
A.~Ryd et al. [BABAR Collaboration],
Proc. Int.~Symp.~on Heavy Flavour Physics 9.

\end{thebibliography}
\end{document}